\documentclass[a4paper]{jpconf}

\usepackage{graphicx}
\usepackage{bm}
\usepackage[usenames]{color}
\usepackage{amsmath}
\usepackage[utf8]{inputenc}
\begin{document}

\makeatletter
\def\@fnsymbol#1{\ensuremath{\ifcase#1\or ^*\or ^\dagger\or \ddagger\or
   \mathsection\or \mathparagraph\or \|\or **\or \dagger\dagger
   \or \ddagger\ddagger \else\@ctrerr\fi}}
\makeatother

\title{Ptychography imaging of the phase vortices in the x-ray beam formed by nanofocusing lenses}
\author{D Dzhigaev$^1$$^,$$^2$, U Lorenz$^1$$^,$\footnote{Present address: Institut für Chemie, Universität Potsdam, Karl-Liebknecht-Straße 24-25, D-14476 Potsdam, Germany}, R Kurta$^1$, F Seiboth$^3$, T Stankevic$^4$, S~Mickevicius$^4$, A Singer$^1$$^,$\footnote{Present address: The University of California, San Diego, La Jolla, CA 92093, USA}, A Shabalin$^1$, O Yefanov$^5$, M~N~Strikhanov$^2$, G Falkenberg$^1$, C G Schroer$^3$, R Feidenhans`l$^4$ and I A Vartanyants$^1$$^,$$^2$}
\address{ $^1$ Deutsches Elektronen-Synchrotron DESY, Notkestra\ss e 85, D-22607 Hamburg, Germany}
\address{ $^2$ National Research Nuclear University, ``MEPhI'', 115409 Moscow, Russia}
\address{ $^3$ Institute for Structural Physics, Technische Universität Dresden, D-01062 Dresden, Germany}
\address{ $^4$ Niels Bohr Institute, University of Copenhagen, DK-1165 Copenhagen, Denmark}
\address{ $^5$ Center for Free-Electron Laser Science CFEL, Notkestra\ss e 85, D-22607 Hamburg, Germany}

\eads{\mailto{dmitry.dzhigaev@desy.de}}

\begin{abstract}
We present the ptychography reconstruction of the x-ray beam formed by nanofocusing lenses (NFLs) containing a number of phase singularities (vortices) in the vicinity of the focal plane. As a test object Siemens star pattern was used with the finest features of 50 nm for ptychography measurements. The extended ptychography iterative engine (ePIE) algorithm was applied to retrieve both complex illumination and object functions from the set of diffraction patterns. The reconstruction revealed the focus size of 91.4$\pm$1.1 nm in horizontal and 70$\pm$0.3 nm in vertical direction at full width at half maximum (FWHM). The complex probe function was propagated along the optical axis of the beam revealing the evolution of the phase singularities.
\end{abstract}

\section{Introduction}
Phase singularities are a common feature of different forms of waves and represent a fundamental topology property of wavefields \cite{Berry}. In 1804 Young has described effects of  interference from different types of obstacles in the path of the light beam \cite{Dennis}. When three or more waves interfere, points of zero intensity could appear. At these positions the phase is undefined (singular), and, in general, all phase values in the interval [0; 2$\pi$] occur around a vortex point, leading to a circulation of the optical energy. Phase singularities were discussed in the terms of dislocations in wave trains in 1974 \cite{Nye} and were first observed in the optics of visible light \cite{Tamm}. Ten years ago phase vortices were observed in x-ray regime using spiral zone plate at 9-keV photon energy \cite{Peele}. Phase singularities could also appear after interaction with the crystal lattice dislocations \cite{Jacques, Taka}. 

One generally describes light as a plane wave, that is an electromagnetic field with a constant phase that extends infinitely and normally to the Poynting vector, $\psi(\textbf{r})$=$\rho(\textbf{r})\exp(i\varphi(\textbf{r}))$, where $\rho(\textbf{r})$ is the real amplitude and $\varphi(\textbf{r})$ is the phase at the position $\textbf{r}$. Vortices can be characterized by the integer number S (positive or negative) that is called a strength or topological charge of the singularity and is determined by S=$({}^1/_{2\pi})\oint\limits_C\mathrm{d}\varphi$, where \textit{C} is any closed path around the vortex point. Dynamics of the vortices include processes of nucleation, annihilation, and propagation in three dimensions \cite{Chen}.

Ptychography, first proposed by Hoppe in the field of electron microscopy more than 40 years ago \cite{Hoppe}, became a well established x-ray microscopy technique during the last decade \cite{Roden, Thib}. Development of the phase retrieval algorithms \cite{Rodenburg} made this approach especially useful for the beam characterization with the use of the test patterns \cite{Schropp}. It allows to reconstruct complex illumination function (probe) and complex object function simultaneously. In this work we apply ptychography technique to reconstruct the wavefield generated by nanofocusing lenses (NFLs) in hard x-ray regime. 

\begin{figure}
\centering
\includegraphics{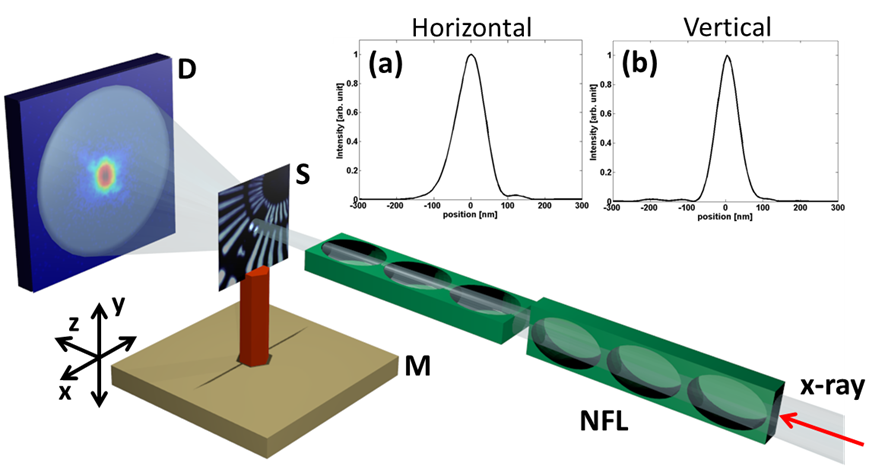}
\caption{Experimental setup. The incoming x-ray beam (red arrow) goes along the Z axis and is focused by a pair of perpendicularly positioned NFLs. The test sample (S) in the form of the Siemens star is mounted on the movable stage (M) and is illuminated at the positions of a raster grid. Diffraction patterns are collected by the detector (D) 2.1 m downstream. Insets (a) and (b) show horizontal and vertical profiles of the intensity of the reconstructed probe function across the central maximum.}
\end{figure}

\section{Experiment}
We performed an experiment at the nanoprobe end station of P06 beamline at PETRA III synchrotron source at DESY \cite{Schroer}. The geometry of the experiment is shown in Fig. 1. Two perpendicularly positioned NFLs based on parabolic compound refractive x-ray lenses (CRLs) were used to obtain a nano-sized focus of the incident x-ray beam with the 15.25 keV energy. These lenses were fabricated by electron beam lithography and deep trench reactive etching. The lenses were produced from silicon, because they can be shaped accurately on the sub-micrometer scale \cite{Boye}. The flux of the beam in the focus was $4\times10^{7}$ photons/sec. Pilatus 300K hybrid-pixel detector (Dectris, Switzerland) with the pixel size of the $172\times172$ $\mu m^2$  was used. Detector was positioned 2.1 m downstream from the sample. A tantalum test sample in the form of the Siemens star, fabricated by nanolithography was mounted on the movable sample stage and positioned in the focal plane. The ptychography scan was performed on a Cartesian grid with 50 nm step size and 41${\times}$41 scan positions, in the horizontal and vertical directions, perpendicular to the optical axis of the beam. The step size corresponds to 58\% of the probe overlap \cite{Bunk}. The acquisition time was 0.3 seconds per scan position. 

\begin{figure}
\centering
\includegraphics{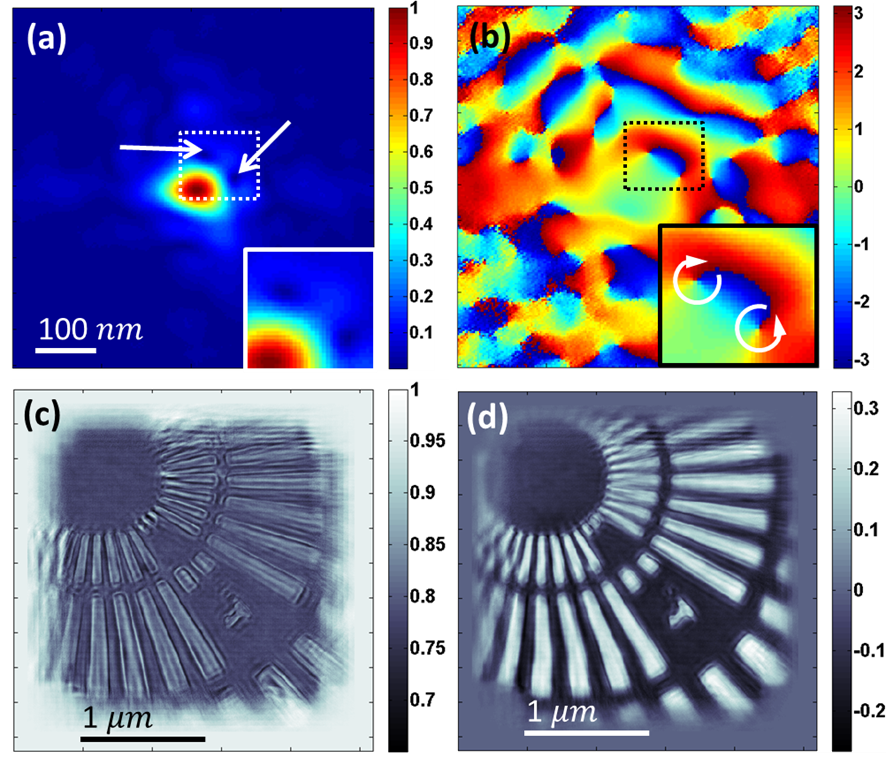}
\caption{Results of ptychography reconstruction at the sample plane. (a) Amplitude of the probe function. White arrows indicate two points of zero intensity corresponding to phase singularities (see inset for an enlarged view). (b) Phase of the probe function. In the inset an enlarged view of two singularities with opposite directions is shown. (c) Amplitude of the object function. (d) Phase of the object function. Smallest resolved lines are 50 nm in size. Color bar in (a) and (c) show normalized values of the amplitude functions and in (b) and (d) values of the phase in radians.}
\end{figure}

\section{Results and discussion}
An extended ptychography iterative engine (ePIE) algorithm \cite{Thibault2} was applied to determine the complex probe (see Fig. 2 (a, b)) and complex object function (see Fig. 2 (c, d)). They were reconstructed from 1681 diffraction patterns. The field of view was about $2\times2$ $\mu m^2$. Reconstruction procedure started from an initial guess of the probe function of a round shape with the uniform positive value inside and zero value outside, and a uniform object with a constant transmissivity without the phase shift. The final result was obtained after 100 iterations. The pixel size in this reconstruction is 6 nm. The smallest detail in the object pattern that was resolved is 20 nm in size (see Fig. 2 (c, d)).

The size of the illumination spot on the sample was obtained by fitting of the intensity distribution by a Gaussian function. It turned out to be 91.4$\pm$1.1 nm in horizontal and 70$\pm$0.3 nm in vertical direction at full width at half maximum (FWHM) (see Fig. 1 (a) and (b)). In the reconstructed images of the probe function (see Fig. 2 (a, b)) one can clearly see a pair of zero amplitude regions and corresponding singularities in the phase. In the inset of Fig. 2 (b) the direction of the phase change is shown by the white arrows. The phase vortices determined by ptychography imaging correspond to the S=\{1,-1\} topological charges.
 
\begin{figure}
\centering
\includegraphics{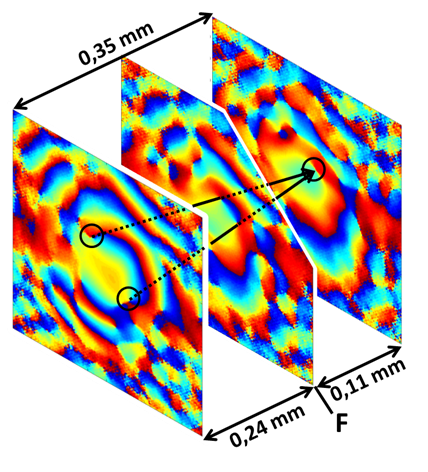}
\caption{Propagation of the wavefield. Three 2D cuts perpendicular to the beam propagation direction are shown: 0.24 mm in front of the focal plane (F), at the focal plane, and 0.11 mm behind the focal plane. At the first position the pair of vortices nucleate, at the last one they annihilate.}
\end{figure}  

The reconstructed complex wave field profile was propagated from 1 mm in front of the focus to 1 mm behind it in the frame of paraxial approximation. The region, where the pair of vortices nucleate and annihilate, is shown in Fig. 3. In three dimensions the wave function $\psi(\textbf{r})$ vanishes at each point of the line, called nodal line. These lines are clearly seen in two dimensional cuts through the propagated wave field along the beam direction (see Fig. 3). The length of the vortex lines is 0.24 mm before and 0.11 mm after the focal plane, giving in total 0.35 mm.

The presence of vortices in the illuminating wavefield may cause degradation of the quality of the phase retrieval procedure. In this case, a number of corrections should be taken into account during reconstruction as shown for high-resolution transmission electron microscopy (TEM) \cite{Allen1} and classical wavefields \cite{Allen2}. Nevertheless, the ptychography technique is very tolerant to imperfections of the probe and the features of the object much smaller than the beam size can be reconstructed as demonstrated in this work.

\section{Conclusions}
We have obtained the ptychography reconstruction of the x-ray field focused by NFLs containing a number of phase singularities (vortices) in the vicinity of the focal plane. Siemens star pattern was used as a test object with the finest features of 50 nm. After inversion procedure with the ePIE algorithm a complex wave field function was obtained. The reconstruction revealed the focus size (FWHM) of 91.4$\pm$1.1 nm in horizontal and 70$\pm$0.3 nm in vertical direction. The pair of vortices closest to the central maximum of the beam with topology charges $\pm$1 were propagated from nucleation to annihilation plane. The length of the nodal line for these singularities was 0.35 mm in total. Appearance of the vortices in the focal region of the nano-focused beam could possibly affect the quality of the phase retrieval procedure and should be taken into account in future work. 

\ack
We acknowledge fruitful discussions and support of this project by E. Weckert, and careful reading of the manuscript by M. Sprung. This work was supported by the EU grant for the project 280773 ``Nanowires for solid state lighting" , the Virtual Institute VH-VI-403 of the Helmholtz Association and by BMBF Proposal 05K10CHG ``Coherent Diffraction Imaging and Scattering of Ultrashort Coherent Pulses with Matter" in the framework of the German-Russian collaboration ``Development and Use of Accelerator-Based Photon Sources".

\section*{References}
\bibliography{bibl}
 
\end{document}